\documentclass[aps,prx,twocolumn,floats,nofootinbib,11pt,tightenlines,superscriptaddress]{revtex4-2}

\usepackage{etoolbox}
\setcitestyle{close}
\usepackage{newtxtext,newtxmath}
\usepackage{mathtools}
\usepackage{wrapfig}
\usepackage{siunitx}
\usepackage{enumerate}
\usepackage{float}
\usepackage{amsmath}
\usepackage{graphicx}
\usepackage[T1]{fontenc}
\usepackage[utf8]{inputenc}
\graphicspath{{images/}}
\usepackage{color}
\usepackage[pdfstartview=FitH,
breaklinks=true,
bookmarksopen=false,
bookmarksnumbered=true,
colorlinks=true,
linkcolor=black,
citecolor=blue,
urlcolor=black,
pdftitle={FoldKit},
pdfauthor={Jonathan Levine},
pdfsubject={}
]{hyperref}

\def\(({\left(}
\def\)){\right)}
\def\[[{\left[}
\def\]]{\right]}

\begin{document}
	
	\title{FoldKit: A Python library for efficient storage and retrieval of co-folding predictions}
	
	\author{Jonathan A. Levine}
	\email{Correspondence: levinej4@mskcc.org}
	\affiliation{Tri-Institutional Program in Computational Biology and Medicine, Weill Cornell Medicine, New York, NY USA}
	\affiliation{Laboratory of Lymphocyte Dynamics, Rockefeller University, New York, NY USA}
	\affiliation{The Halvorsen Center for Computational Oncology, Department of Epidemiology and Biostatistics, Memorial Sloan Kettering Cancer Center, New York, NY USA}
	\author{Melissa Pathil}
	\affiliation{The Halvorsen Center for Computational Oncology, Department of Epidemiology and Biostatistics, Memorial Sloan Kettering Cancer Center, New York, NY USA}
	\affiliation{Computational and Systems Biology Program, Sloan Kettering Institute, Memorial Sloan Kettering Cancer Center, New York, NY, USA}
	\affiliation{Physiology, Biophysics \& Systems Biology, Weill Cornell Medicine, Weill Cornell Medical College, New York, NY USA}
	\author{Samuel Nitz}
	\affiliation{Tri-Institutional Program in Computational Biology and Medicine, Weill Cornell Medicine, New York, NY USA}
	\affiliation{Laboratory of Host-Pathogen Biology, The Rockefeller University, New York, NY 10065, USA}
	\author{Olga Lyudovyk}
	\affiliation{The Halvorsen Center for Computational Oncology, Department of Epidemiology and Biostatistics, Memorial Sloan Kettering Cancer Center, New York, NY USA}
	\author{Benjamin D. Greenbaum}
	\affiliation{The Halvorsen Center for Computational Oncology, Department of Epidemiology and Biostatistics, Memorial Sloan Kettering Cancer Center, New York, NY USA}
	\affiliation{Physiology, Biophysics \& Systems Biology, Weill Cornell Medicine, Weill Cornell Medical College, New York, NY USA}
	\affiliation{The Olayan Center for Cancer Vaccines, Memorial Sloan Kettering Cancer Center, New York, NY USA}
	
	\date{\today}
	
	\def\thefootnote{\arabic{footnote}}
	
	\begin{abstract}
		
		AlphaFold 3 (AF3) enables structure prediction of biomolecular complexes through co-folding multiple interacting molecules, making it increasingly useful for \textit{de novo} protein design and for large-scale studies of protein--protein, protein--peptide, and other biomolecular interactions. However, systematic co-folding experiments can produce large volumes of output data, particularly when multiple random seeds and samples are generated for each input complex. We introduce FoldKit, a Python package for efficient storage and analysis of large-scale AF3 co-folding results. FoldKit converts raw AF3 outputs into a compact, structured representation while preserving the metadata needed for downstream analysis. The FoldKit Python library provides convenient programmatic access to global, single chain, and interface confidence metrics such as pLDDT, pTM, ipTM, ipAE, and ipSAE, as well as an ensemble-level interface for accessing and aggregating these metrics for a single input across multiple seeds and samples. We benchmark FoldKit on three types of AF3 co-folding datasets: (i) a protein design campaign with 2 chains per input, (ii) a TCR-pMHC dataset with 4 chains per input, and (iii) a pooled-AF3 protein--protein interaction dataset with up to 22 chains per input. We find that FoldKit reduces storage requirements by approximately 5--15-fold compared to native AF3 outputs, depending on dataset composition, while maintaining direct programmatic access to individual predictions, ensembles, and confidence metrics. By reducing storage requirements and facilitating programmatic access to relevant outputs, FoldKit facilitates large-scale computational studies of biomolecular interactions. FoldKit is available from PyPI and can be installed using pip.
	\end{abstract}
	
	\maketitle
	
	\section*{INTRODUCTION}
	In recent years, co-folding has become an increasingly popular approach for a variety of tasks in computational biology. Multiple molecules are folded together using a structure prediction model, such as AlphaFold 3 (AF3)~\cite{Abramson2024-yu}, and the resulting structures and confidence metrics can then be used to assess putative interactions between the input molecules, which are represented as separate chains within a single complex. Co-folding has become an important component of \textit{de novo} binder design pipelines, where predictions of a designed protein sequence in complex with its target can effectively prioritize candidates prior to validation, increasing the experimental success rate~\cite{Bennett2023-vg,Watson2023-zt,Overath2025-yc,Cotet2025-fv,Pacesa2025-ol,Chow2025-me,litefold2026litemol1}. These methods have been subsequently adapted to study the binding of larger, native protein complexes, such as antibodies~\cite{Abramson2024-yu,Yin,yin2026benchmarking} or T cell receptors (TCRs)~\cite{lyudovyk2026ensembles,woods2026general,yin2026benchmarking,Richardson2026-xe}, or for more general protein--protein interaction predictions~\cite{humphreys2021computed,bryant2022improved,burke2023towards,guzman2024alphacrv,genz2025assessing,pei2026structural}. More recently, a ``pooled'' version of AlphaFold3-based co-folding was developed, where as many as 25 proteins are folded together at once, allowing numerous potential pairwise interactions to be screened simultaneously in one prediction~\cite{todor2026predicting}.
	
	\begin{figure*}[t!]
		\includegraphics[width=\textwidth]{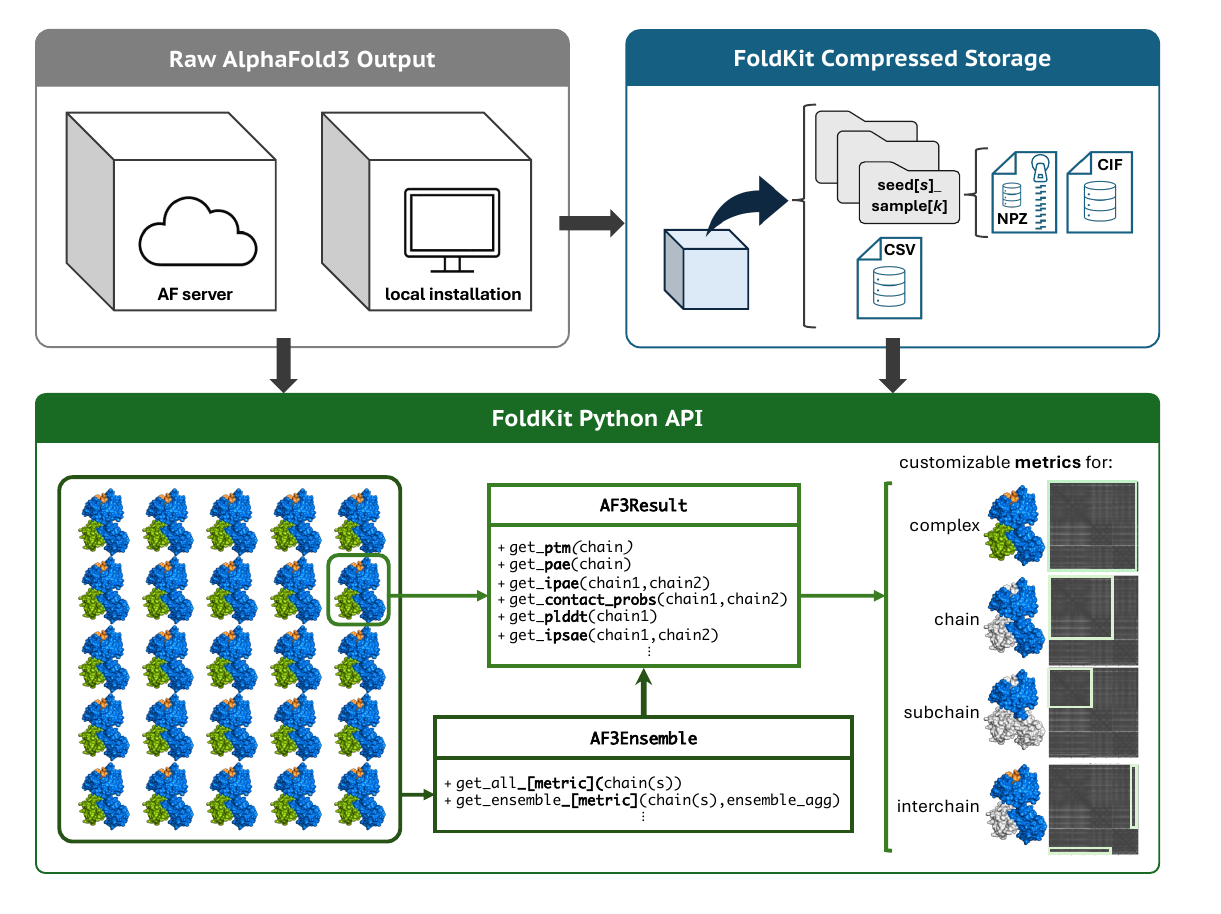}
		\caption{{\bf The FoldKit Package}\\ FoldKit can process AlphaFold 3 results from either the AlphaFold 3 Server or from a local installation. The FoldKit command-line interface (CLI) exports AF3 results into a compressed format using NPZ files. The FoldKit Python API (application programming interface) can read AF3 results from either the raw output or the FoldKit compressed storage. The API enables users to programmatically access customizable metrics for the global complex, individual chains, subchains, or pairs of chains. The AF3Ensemble API enables convenient access to predictions across ensembles of multiple seeds and samples for the same input, where users can access metrics for the entire ensemble or choose how to aggregate them by passing in a custom function.}
		\label{fig:api}
	\end{figure*}
	
	These methods utilize a variety of outputs from the AlphaFold pipeline, including the predicted local distance difference test (pLDDT), the predicted alignment error (PAE), and the predicted template-modeling score (pTM). Usually the most informative metrics for binding are computed by aggregating and/or transforming these confidence values over the interchain interfaces (in the typical notation they get prefixed with an ``i''), like ipAE, ipTM, or ipSAE~\cite{dunbrack2025res}.
	
	\indent Confidence values are output by the AlphaFold 3 pipeline or the AlphaFold 3 Server in large JSON (JavaScript Object Notation) files alongside the predicted structure file. JSON is a text-based format for storing structured data, in which numerical values and other objects are represented as human-readable text. While these JSON files contain all of the information needed for downstream analyses, their storage format leads to two major pain points for scientists:
	
	\begin{enumerate}
		\item Extracting specific interchain or subchain metrics from these files can require substantial parsing and custom code, particularly when analyzing many predictions or heterogeneous input complexes.
		
		\item The data is stored inefficiently, as numerical values are represented as text, and the storage needed for directories of co-folding results for large datasets can become prohibitively expensive.
	\end{enumerate}

	To solve these issues, we developed \textbf{FoldKit}.
	
	\section*{FoldKit}
	The FoldKit Python API (application programming interface) provides a convenient interface for loading AF3 co-folding results generated either through the AlphaFold 3 Server or local installations, and for extracting and aggregating confidence metrics from the results \textbf{(Fig.~\ref{fig:api})}. The library allows for customization of features, including user-defined aggregation functions, extraction of specific chain/interchain data, and extraction of data for partial chain subsequences. There is a programmatic interface for extracting features from an ensemble of predictions for a single sequence input using multiple random seeds and samples, which has been shown to help with downstream classification tasks~\cite{lyudovyk2026ensembles}.\\
	\indent The FoldKit CLI (command-line interface) enables the export of raw AF3 JSON files into a compressed, structured representation using .npz files, and provides built-in integration between the exported data and the Python API layer for accessing ensembles and individual results.
	
	\begin{figure*}[t!]
		\includegraphics[width=\textwidth]{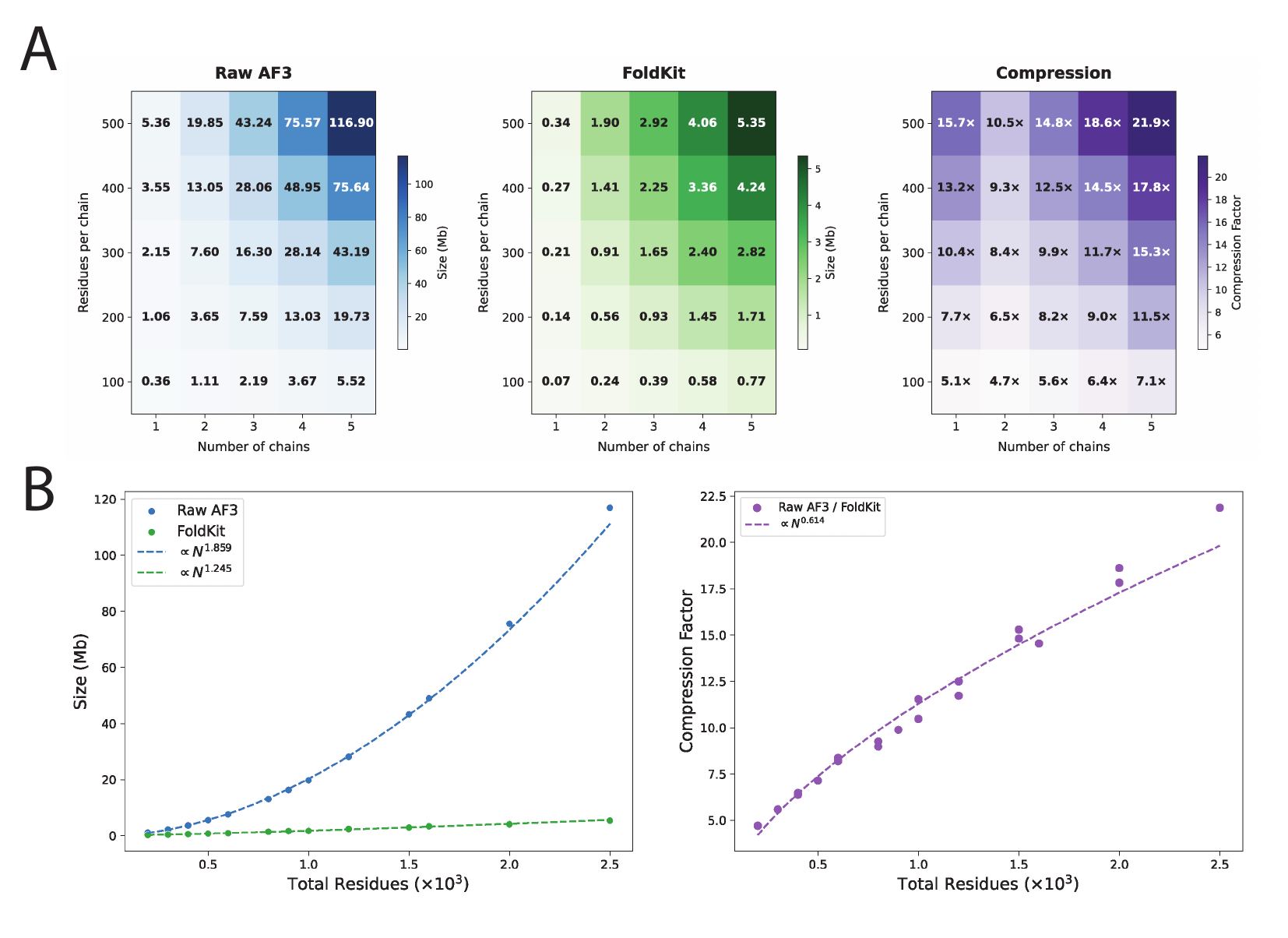}
		\caption{{\bf FoldKit substantially reduces storage requirements on a synthetic co-folding dataset}\\
			\textbf{A:} Size (Mb) of output directories for a single AF3 prediction (1 seed, 1 sample, $N_{\mathrm{Ensemble}}=1$) as a function of the number of chains and number of residues-per-chain for raw AF3 (left) and after FoldKit export (center).  Right: The compression factor is defined as the ratio of raw AF3 size to FoldKit exported size.
			\textbf{B:} Left: Storage size (Mb) scaling-laws for raw AF3 (blue) and FoldKit (green). Raw AF3 is fit as a power-law with exponent $1.859$ and FoldKit as a power-law with exponent $1.245$. Right: Compression factor scaling with number of residues is fit as a power-law with the ratio of exponents from the individual power laws $(1.859-1.245=0.614)$. Total residues is calculated as $N_{\mathrm{chains}}*N_{\mathrm{res. per ~chain}}$}.
		\label{fig:synthetic}
	\end{figure*}

	\section*{FoldKit Compression on Synthetic Data}
	To evaluate the compression efficiency of FoldKit, we ran AlphaFold 3 with 1 seed and 1 sample on a synthetically generated dataset in which we varied both the number of chains and the number of residues per chain (\textbf{Fig.~\ref{fig:synthetic}}). We compared the storage requirements of the raw AF3 output directories with those of the corresponding FoldKit exports. FoldKit reduced the storage requirements of individual AF3 outputs by approximately 5- to 20-fold (\textbf{Fig.~\ref{fig:synthetic}A)}. The size of raw AF3 outputs scaled with the total number of residues as a power law with an exponent of 1.859, consistent with the dominant contribution of the $O(N^2)$ residue-pair confidence matrices to storage requirements (\textbf{Fig.~\ref{fig:synthetic}B, left)}. In contrast, FoldKit outputs exhibited substantially weaker scaling, with a power-law exponent of 1.245, reflecting the use of efficient matrix storage and removal of redundant data (\textbf{Fig.~\ref{fig:synthetic}B, right)}. Consequently, the FoldKit compression factor (raw AF3 size / FoldKit size) increased with complex size, as expected from the difference in scaling exponents. Thus, larger AF3 complexes derive progressively greater storage savings from FoldKit.

			\begin{figure*}[t!]
	\includegraphics[width=\textwidth]{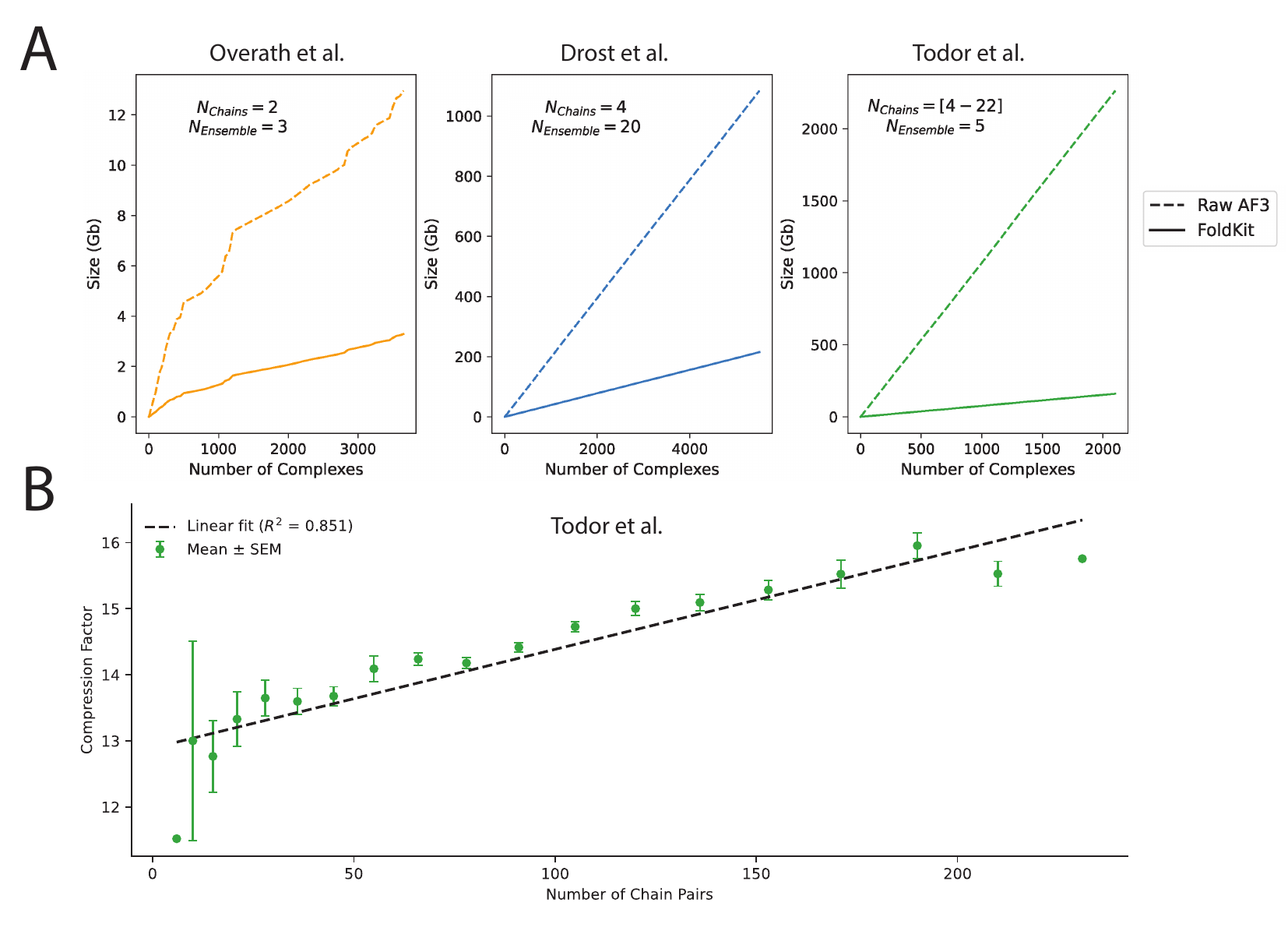}
	\caption{{\bf FoldKit substantially reduces storage requirements on real AF3 co-folding data}\\
		\textbf{A:} Cumulative storage size (Gb) as a function of the number of input complexes for raw AF3 output (dashed) versus FoldKit output (solid), shown for the Overath et al. ($N_{\mathrm{chains}}=2$, $N_{\mathrm{Ensemble}}=3$), Drost et al. ($N_{\mathrm{chains}}=4$, $N_{\mathrm{Ensemble}}=20$), and Todor et al. ($N_{\mathrm{chains}}=4\text{--}22$, $N_{\mathrm{Ensemble}}=5$) datasets. In all three datasets, both raw AF3 and FoldKit storage scale linearly with the number of complexes, but with a substantially shallower slope for FoldKit.
		\textbf{B:} Compression factor (ratio of raw AF3 to FoldKit output size) as a function of the number of chain pairs, $\binom{N_{\mathrm{chains}}}{2}$, for the Todor et al. dataset. Points show the mean $\pm$ SEM compression factor across complexes with a given number of chain pairs; the dashed line shows a linear fit ($R^2=0.851$).}
	\label{fig:compression}
\end{figure*}
	\section*{FoldKit Compression on Public Data}
To evaluate the utility of FoldKit on real-world data, we analyzed three existing AF3 co-folding datasets spanning a range of input sizes, ensemble sizes, and numbers of input chains:

\begin{enumerate}[(i)]
	\item the Overath et al. data~\cite{Overath2025-yc,overath_2025_15722219} with $\approx 3700$ input complexes consisting of binder-target pairs, each run with 1 AF3 seed and 3 AF3 samples ($N_{Ensemble}=3$),
	\item the Drost et al. data sampled from~\cite{drost2024predicting} (co-folded as part of Lyudovyk et al.~\cite{lyudovyk2026ensembles}), comprising $\approx 5500$ TCR-pMHC complexes with 4 chains each (TCR$\alpha$, TCR$\beta$, MHC, peptide), each run with 4 AF3 seeds and 5 AF3 samples ($N_{Ensemble}=20$),
	\item the Todor et al. data sampled from~\cite{todor2026predicting,todor_2025_15499631}, containing $\approx 2100$ pooled complexes with the number of proteins (i.e. chains) per complex normally distributed over 4--22 ($\mu=12.8$, $\sigma=2.6$), run on the AlphaFold 3 Server with 1 seed and 5 samples ($N_{Ensemble}=5$).
\end{enumerate}

Across all three datasets, exporting AF3 results with FoldKit substantially reduced storage requirements \textbf{(Fig.~\ref{fig:compression}A)}. For each dataset, both raw AF3 and FoldKit storage requirements increased approximately linearly with the number of input complexes, as expected when storing an increasing number of independent results. However, the FoldKit outputs exhibited substantially shallower scaling, resulting in progressively larger storage savings as the number of complexes increased \textbf{(Fig.~\ref{fig:compression}A)}. The magnitude of these savings also depended on the size and complexity of the individual complexes, consistent with the scaling behavior observed in the synthetic dataset.

We next examined how the number of input chains influenced FoldKit's compression using the Todor et al. dataset. Increasing the number of chains introduces additional pairwise confidence information because AF3 generates pairwise predictions between each pair of input chains. The number of chain pairs increases as $\binom{N_{\mathrm{chains}}}{2}$, making pairwise confidence information an increasingly important component of the raw output as the number of chains grows. Consistent with this expectation, the compression factor increased approximately linearly with the number of chain pairs, $\binom{N_{\mathrm{chains}}}{2}$ (\textbf{Fig.~\ref{fig:compression}B}). This provides further evidence that the efficient representation of pairwise confidence information is a major source of FoldKit's storage savings, though the exact performance depends on the number of residues in each chain, which is highly variable in this dataset. Thus, FoldKit provides particularly substantial storage benefits for large multi-chain AF3 jobs, such as those generated by pooled-AF3 approaches.

	\section*{DISCUSSION}
	FoldKit addresses a practical bottleneck in large-scale AF3 co-folding studies. As co-folding is increasingly applied to protein design, TCR-pMHC or antibody specificity prediction, and large pooled interaction screens, the raw JSON outputs from AF3 quickly become unwieldy, both in terms of storage footprint and the effort required to extract relevant confidence metrics. By converting these outputs into a compact, structured representation, FoldKit reduces storage costs by roughly an order of magnitude while preserving direct programmatic access to per-residue and per-interface confidence metrics, individual predictions, and ensembles across seeds and samples.
	
	The benefits of FoldKit scale with the complexity of the input, since the number of pairwise confidence matrices grows quadratically with the number of chains. This makes FoldKit particularly useful for emerging pooled co-folding approaches, where dozens of chains may be folded together in a single job and the corresponding pairwise interface data would otherwise dominate storage requirements. Similarly, the version of ipSAE included in FoldKit is rewritten to use numpy's broadcasting feature, speeding up calculation for AlphaFold outputs with many chains.
	
	There are several limitations and directions for future work. First, FoldKit currently supports AF3 outputs (both local and server-based runs); extending support to other co-folding tools would broaden its utility as the co-folding ecosystem diversifies. Second, our benchmarks were conducted on three datasets spanning binder design, TCR-pMHC, and pooled interaction prediction; testing on additional dataset types, including complexes with non-protein chains (e.g., nucleic acids~\cite{meng2026precise} or ligands~\cite{vskrinjar2026evaluating}), would help characterize how well FoldKit's compression generalizes beyond protein-only inputs. Finally, while FoldKit focuses on efficient storage and retrieval, natural extensions include tighter integration with downstream modeling and visualization tools, allowing FoldKit exports to be used directly as inputs to classifiers or structural analysis pipelines without additional conversion steps.
	
	Overall, FoldKit provides a lightweight, general-purpose solution to a storage and accessibility problem that will only grow as co-folding is applied more broadly across structural biology.
	
	\section*{CODE AVAILABILITY}
	The code for FoldKit can be found at \url{https://github.com/jonlevi/foldkit}. FoldKit can be installed using pip, as described in the git repo. Full documentation for FoldKit can be found at \url{https://jonlevi.github.io/foldkit/}.
	
	\section*{ACKNOWLEDGMENTS}
	This work utilized resources from the High-Performance Computing (HPC) Group at Memorial Sloan Kettering Cancer Center. This work was in part supported by the Robert J. Kleberg, Jr. and Helen C. Kleberg Foundation. JAL is supported by NIAID Ruth L. Kirschstein National Research Service
	Award for Predoctoral Fellows 1F31AI200147-01.
	
	\section*{References}
	\def\bibsection{}
	\bibliographystyle{naturemag}
	\bibliography{library}
	
\end{document}